\documentclass[a4paper,11pt]{article}
\usepackage{pos}
\usepackage{bbold}

\let\oldexp\exp
\renewcommand{\exp}[1]{\oldexp \left\{ #1\right\}}

\newcommand{\mD}{\mathcal{D}}

\newcommand{\secref}[1]{Sec.~\ref{#1}}

\newcommand{\figref}[1]{Fig.~\ref{#1}}

\renewcommand{\eqref}[1]{(\ref{#1})}
\newcommand{\Eqref}[1]{Eq.~(\ref{#1})}

\title{Hybrid Monte Carlo for the extended Kane-Mele-Hubbard model}

\author*[a]{Finn L. Temmen}
\author[a,b]{Thomas Luu}
\author[b]{Johann Ostmeyer}
\author[a]{Lin Wang}

\affiliation[a]{Institute for Advanced Simulation (IAS-4),\\
Forschungszentrum Jülich, Germany}

\affiliation[b]{Helmholtz-Institut für Strahlen- und Kernphysik and Bethe Center for Theoretical Physics,\\
Rheinische Friedrich-Wilhelms-Universität Bonn, Germany}

\emailAdd{f.temmen@fz-juelich.de}
\emailAdd{t.luu@fz-juelich.de}
\emailAdd{ostmeyer@hiskp.uni-bonn.de}
\emailAdd{l.wang@fz-juelich.de}

\abstract{
	The Hubbard model offers a powerful minimal description of strongly correlated electrons in molecules and materials, yet it neglects effects observed in nature that give rise to a much richer variety of physics.
	For instance, including nonlocal interactions introduces a competition between strongly correlated phases, while incorporating spin-orbit coupling (SOC) can lead to the emergence of exotic topological phases.
	Seeking a more realistic platform for simulating graphene and, potentially, related materials such as transition metal dichalcogenides (TMDs), we therefore consider the extended Kane-Mele-Hubbard model, which features a Kane-Mele SOC term and nearest-neighbor interactions.
	
	Simulating this model is particularly challenging once the nonlocal interactions become dominant over the on-site Hubbard interaction, as a severe sign problem arises already at half filling, leaving this regime largely unexplored.
	In this work, we develop continuous auxiliary field formulations of the model, suitable for treatment with the Hybrid Monte Carlo (HMC) algorithm.
	Using these formulations, we characterize the sign problem across the model's parameter space and establish where simulations are feasible, as a first step toward the full phase diagram.
}

\FullConference{The 43rd International Symposium on Lattice Field Theory (Lattice 2026)\\
July 26 to August 1, 2026\\
University of Maryland, College Park, USA\\}

\begin{document}
	\maketitle
	
	\section{Introduction}
	\label{sec:introduction}
		The Hubbard model is a widely used minimal effective description of correlated electrons in molecules and materials.
		Combining a tight-binding hopping term with a purely on-site interaction, it captures essential phenomena observed in nature, such as the transition from a semimetal to an antiferromagnetic Mott insulator \cite{ostmeyer_semimetalmott_2020,ostmeyer_antiferromagnetic_2021}.
		Its simplicity has made it particularly popular for the description of low-dimensional carbon-based systems, such as perylene \cite{rodekamp_single-particle_2025}, carbon nanotubes \cite{Luu:2015gpl}, and most prominently graphene \cite{Brower:2011nqf}.

		However, despite its success, the Hubbard model neglects a number of effects present in real materials, and extending it in different directions gives rise to a considerably richer variety of physics.
		For instance, generalizing the on-site interaction to nonlocal interactions, as in the Coulomb-Hubbard model \cite{buividovich_hybrid_2018,hohenadler_phase_2014} or the extended Hubbard model \cite{kennedy_extended_2025}, introduces a competition between charge density wave (CDW) and spin density wave (SDW) order that is absent in the standalone Hubbard model.
		Separately, adding a spin-orbit coupling (SOC) term, as in the Kane-Mele model \cite{kane_quantum_2005,kane_z_2005} for example, can drive the system into a topological phase and give rise to quantum spin Hall physics.

		In an effort to provide a more realistic framework for the simulation of such systems, we therefore consider the extended Kane-Mele-Hubbard model, which comprises both extensions.
		Simulating this model is particularly challenging once the nonlocal interaction becomes dominant over the on-site interaction, as a severe sign problem arises already at half filling.
		Therefore, the regime has largely been avoided in the literature, and the SOC term is expected to compound this difficulty further by introducing a complex term into the connectivity matrix.
		
		In this work, we develop continuous auxiliary field formulations of the extended Kane-Mele-Hubbard model suitable for treatment with the Hybrid Monte Carlo (HMC) algorithm \cite{HMC}.
		We show that the freedom in the Hubbard-Stratonovich transformation gives rise to several alternative formulations, and exploit it to reach the regime of dominant nonlocal interactions and establish where simulations remain feasible.
	
	\section{Model}
	\label{sec:model}
		We consider the extended Kane-Mele-Hubbard model on a two-dimensional honeycomb lattice with periodic boundary conditions.
		The lattice is bipartite with $K=3$ nearest neighbors per site, and a linear system size $L$ corresponds to $N_x = 2L^2$ spatial sites.
		We formulate the model in the particle/hole basis, in which the fermionic ladder operators $c_{x\uparrow}^{\vphantom{\dagger}},c_{x\uparrow}^{\dagger}$ ($c_{x\downarrow}^{\vphantom{\dagger}},c_{x\downarrow}^{\dagger}$) for spin-$\uparrow$ (spin-$\downarrow$) electrons are replaced by the particle/hole operators
		\begin{equation}
			\label{eq:particle_hole_trafo}
			a_{x}^\dagger \equiv c_{x\uparrow}^{\dagger}; \quad
			a_{x}^{\vphantom{\dagger}} \equiv c_{x\uparrow}^{\vphantom{\dagger}}; \quad
			b_{x}^\dagger \equiv c_{x\downarrow}^{\vphantom{\dagger}}; \quad
			b_{x}^{\vphantom{\dagger}} \equiv c_{x\downarrow}^{\dagger}.
		\end{equation}
		In terms of these operators, their density operators $q_{x} \equiv a_{x}^\dagger a_{x}^{\vphantom{\dagger}}$ and $\tilde{q}_{x} \equiv b_{x}^\dagger b_{x}^{\vphantom{\dagger}}$, and the shorthand $Q_x \equiv q_x - \tilde{q}_{x}$, the Hamiltonian reads
		\begin{equation}
			\label{eq:ekmh_hamiltonian}
			\begin{split}
				H 	=  
				&-\kappa \sum_{\langle x,y \rangle}
					\left(
						a_{x}^\dagger a_{y}^{\vphantom{\dagger}}
						- b_{x}^\dagger b_{y}^{\vphantom{\dagger}}
					\right) 
				+ \mathrm{h.c.}
				+ i\lambda_{\mathrm{SO}} \sum_{\langle\langle x,y \rangle\rangle}  
					\nu_{xy} \left(
						a_{x}^\dagger a_{y}^{\vphantom{\dagger}}
						- b_{x}^\dagger b_{y}^{\vphantom{\dagger}} 
					\right) + \mathrm{h.c.}\\
				&+ U \sum_x q_x (1-\tilde{q}_x)
				+ V \sum_{\langle x,y \rangle} (Q_x + 1)(Q_y + 1)
				+ (\mu -\frac{U}{2} - KV) \sum_x Q_x.	
			\end{split}
		\end{equation}
		The model comprises the usual terms of the Hubbard model with nearest-neighbor hopping ($\kappa$), a chemical potential ($\mu$), and an on-site interaction ($U$).
		The chemical potential has been shifted such that the system is at half filling for $\mu=0$.
		The Hubbard model's local $U$ interaction is extended by a nearest-neighbor interaction ($V$), giving rise to the name \emph{extended} Hubbard model.
		Finally, a Kane-Mele SOC ($\lambda_{\mathrm{SO}}$) introduces a direction-dependent complex next-nearest-neighbor hopping that is encoded in the connectivity matrix $\nu_{xy} = \pm 1$.
		This matrix takes the value $+1$ ($-1$) when hopping from $x$ to $y$ via the shared nearest neighbor involves a counterclockwise (clockwise) turn.
		Physically, the SOC describes a relativistic effect in which an electron moving through an electrostatic potential experiences an effective magnetic field in its rest frame, to which its intrinsic magnetic moment couples.
		Throughout this work, all physical quantities are given in units of the hopping amplitude $\kappa$.

	\section{Auxiliary field formulation}
	\label{sec:formulation}
		Simulating the model with HMC requires a path integral formulation in which the many-body interactions are replaced by fermion bilinears coupled to continuous auxiliary fields.
		We obtain such a formulation by completing the interaction terms of \Eqref{eq:ekmh_hamiltonian} into squares, performing a Suzuki-Trotter decomposition of the partition function, and decoupling the squares by means of a Hubbard-Stratonovich (HS) transformation.
		The completion of the squares leaves a freedom of choice that carries over to the type of HS transformation, and the resulting formulations differ substantially in their sampling properties and regions of applicability.
		In the following, we sketch this derivation and discuss the two auxiliary field formulations used throughout this work.

		\subsection{Hubbard-Stratonovich transformation}
		\label{sec:HS_trafo}
			In general, an HS transformation corresponds to a Gaussian integral identity
			\begin{equation}
			\label{eq:general_HS_transformation}
				\exp{\pm \frac{|\alpha|}{2} \mathcal{O}^2} 
				\propto \int_{-\infty}^{\infty} d\phi
					\exp{
						-\frac{\phi^2}{2|\alpha|}
						- \sqrt{\pm 1} \phi \mathcal{O}
					}
			\end{equation}
			that linearizes the square of an operator $\mathcal{O}$ at the cost of introducing an auxiliary field $\phi$ to be integrated over.
			For Hubbard-like models, applying this identity to the density-density interactions leaves only fermion bilinears in the exponent, which may then be integrated out to yield the familiar fermion determinants.

			Equation~\eqref{eq:general_HS_transformation} admits two different variants, depending on the sign of the exponent on the left-hand side.
			A positive argument couples the auxiliary field to the operator with a real prefactor of $1$, whereas a negative argument requires a complex prefactor of $i$.
			We refer to the resulting transformations as real and imaginary HS transformations, respectively.
			Since the interaction terms can be completed into squares of either sign, as shown below, both transformations are available for the same interaction, but they give rise to auxiliary field formulations with significantly different sampling properties.
			For the model at hand, we will exploit this ambiguity in two ways.
			The first yields a formulation that is free of the sign problem at half filling, but restricted to the region $U\geq KV$.
			The second is not protected from the sign problem anywhere, but it remains applicable for $U< KV$ and, as we show in \secref{sec:results}, still permits feasible simulations.

		\subsection{Completing the squares}
		\label{sec:completing_squares}
			To see how the ambiguity in the formulations arises, we return to the interaction terms of \Eqref{eq:ekmh_hamiltonian}, which have to be completed into squares before the HS transformation \eqref{eq:general_HS_transformation} can be applied.
			Starting with the $V$ interaction, we expand the product and absorb the term linear in $Q_x$ into the chemical potential, leaving only the many-body contribution up to an irrelevant constant.
			Completing the square then yields
			\begin{equation}
				V\sum_{\langle x,y\rangle} Q_xQ_y 
					= \sigma_V \frac{V}{2}\sum_{\langle x,y\rangle} \left(Q_x + \sigma_V Q_y\right)^2
					-\sigma_V \frac{KV}{2} \sum_{x} \left(q_x + \tilde{q}_x\right)
					+\sigma_V KV \sum_x q_x \tilde{q}_x, 
			\end{equation}
			where $\sigma_V= \pm 1$ can be freely chosen.
			In addition to the squared operator, this generates a bilinear term that is absorbed into the chemical potential, as well as a shift of $\sigma_V KV$ to the on-site interaction strength.
			Taking the latter into account and applying the same procedure to the remaining on-site interaction term yields
			\begin{equation}
				-(U-\sigma_V KV) \sum_x q_x\tilde{q}_x = - \sigma_U \frac{U-\sigma_V KV}{2} \sum_x
					\left(
						q_x + \sigma_U \tilde{q}_x
					\right)^2
					+ \sigma_U \frac{U-\sigma_V KV}{2} \sum_x \left(q_x + \tilde{q}_x\right)
			\end{equation}
			and introduces a second sign $\sigma_U = \pm 1$ that may again be chosen freely.

			Both interactions are now squares of operators, and we group the Hamiltonian accordingly into
			\begin{equation}
				H = H_{2} + H_V + H_U, 
			\end{equation}
			where
			\begin{equation}
				\begin{split}
					H_V &= c_V \sum_{\langle x,y\rangle} \left(Q_x + \sigma_V Q_y\right)^2, 
						\qquad c_V \equiv \sigma_V \frac{V}{2},\\
					H_U &= c_U \sum_x
					\left(
						q_x + \sigma_U \tilde{q}_x
					\right)^2, \qquad c_U \equiv - \sigma_U \frac{1}{2} (U-\sigma_V KV),
				\end{split}
			\end{equation}
			and $H_{2}$ collects the remaining bilinear terms.
			The signs of $c_V$ and $c_U$ determine which of the two HS transformations is required and can always be set through the choice of $\sigma_U$ and $\sigma_V$.
			In the case of $\sigma_V = +1$, the type of transformation for the on-site interaction additionally depends on the relative strength of $KV$ and $U$.
			This dependence on $\sigma_V$ and $\sigma_U$ is inherited by all following quantities, but we suppress it in the notation for clarity.

		\subsection{Path integral}
		\label{sec:path_integral}
			Interpreting the inverse temperature $\beta$ as an extent in Euclidean time and discretizing it into $N_t$ time slices of size $\Delta_t = \beta /N_t$, we can split the exponential using a second-order Suzuki-Trotter decomposition.
			Since $H_V$ and $H_U$ contain only number operators and therefore commute, the half steps of adjacent time slices may be merged using the cyclicity of the trace, such that
			\begin{equation}
				\label{eq:Z_trotter}
				Z = \mathrm{tr}\left[ e^{-\beta (H_{2} + H_V + H_U)} \right]
				=  \mathrm{tr}
				\left[ \prod_{t=1}^{N_t} 
					\left(
						e^{-\Delta_t H_{2}}e^{-\Delta_t H_V}e^{-\Delta_t H_U} 
					\right)		
				\right] + \mathcal{O}\left(\Delta_t^2\right).
			\end{equation}

			Applying \Eqref{eq:general_HS_transformation} to $H_V$ and $H_U$ introduces an auxiliary field $\phi$ for every bond $\langle x,y\rangle$ and a field $\chi$ on every site $x$ of each time slice $t$, leaving only fermion bilinears in the exponents.
			Integrating out the fermions then yields the path integral representation
			\begin{equation}
				\label{eq:final_Z}
				Z = \int \mD\phi \mD \chi ~
					e^{-S[\phi,\chi]} + \mathcal{O}\left(\Delta_t^2\right)
					\quad \text{with} \quad 
					S = S_{\mathrm{B}}[\phi,\chi] + S_{\mathrm{F}}[\phi,\chi],
			\end{equation}
			where $\mD\phi \equiv \prod_{t}\prod_{\langle x,y\rangle} d\phi_{t\langle x,y\rangle}$ and $\mD\chi \equiv \prod_{t}\prod_{x} d\chi_{tx}$ are the high-dimensional integration measures and $S_{\mathrm{B}}$ collects the purely Gaussian auxiliary field contributions from \Eqref{eq:general_HS_transformation}.
			The fermionic contribution $S_{\mathrm{F}}[\phi,\chi] = - \log\left( \det M[\phi,\chi] \det \bar{M}[\phi,\chi]\right)$ is given in terms of the particle and hole fermion matrices
			\begin{equation}
				\label{eq:fermion_matrices}
				\begin{split}
					M[\phi,\chi] = \mathbb{1} + \prod_{t=1}^{N_t} \left( e^{A} e^{B^{(t)}[\phi,\chi]} \right)
					\quad \text{and} \quad
					\bar{M}[\phi,\chi] = \mathbb{1} + \prod_{t=1}^{N_t} \left(e^{\bar{A}} e^{\bar{B}^{(t)}[\phi,\chi]}\right).
				\end{split}
			\end{equation}
			Here, the matrices $A,\bar{A}$ are obtained from the bilinear part $H_2$ and carry no time dependence, while the diagonal matrices $B^{(t)}[\phi,\chi], \bar{B}^{(t)}[\phi,\chi]$ contain the contributions coupling the auxiliary fields to the density operators on each time slice.
			We denote expectation values of an operator $\mathcal{O}$ with respect to an action $S$ by $\langle \mathcal{O} \rangle_S$.

		\subsection{The two formulations}
		\label{sec:two_formulations}
			We first consider the sign choice $\sigma_V = +1$, leading to an imaginary HS transformation to decouple the $V$ interaction.
			At the same time, this choice affects the decoupling of the on-site interaction, which enters through $c_U = - \sigma_U \frac{1}{2} (U-\sigma_V KV)$, whose sign is then determined by both $\sigma_U$ and the relative strength of $U$ and $KV$.
			With $\sigma_U = -1$, the on-site interaction is decoupled by a second imaginary transformation for $U \geq KV$.
			The resulting \emph{imaginary field} formulation can be shown to have a positive weight at half filling, which is preserved by the Kane-Mele SOC term \cite{zheng_particle-hole_2011}.
			Once the nonlocal interaction dominates, however, this transformation turns real, the weight loses its positivity, and we encounter a prohibitive sign problem.
			We therefore adopt the \emph{real field} formulation for $KV > U$, given by the sign choice $\sigma_V = -1$ and $\sigma_U = +1$, which decouples both interactions with real HS transformations.
			Since the corresponding coefficient $c_U = -\frac{1}{2}\left( U + KV\right)$ is negative for all positive $U$ and $V$, this formulation remains available throughout the parameter range.
			Moreover, while it does not restore positivity of the overall weight at half filling, the resulting sign problem is mild enough to permit feasible simulations for $KV>U$, as we demonstrate in \secref{sec:results}.

	\section{Results}
	\label{sec:results}
		Our aim is to establish where simulations of the extended Kane-Mele-Hubbard model are feasible, as a first step toward mapping out its phase diagram.
		To this end, we use the framework of the previous section to scan the parameter space and quantify the severity of the sign problem through the average phase
		\begin{equation}
			\label{eq:average_phase}
			\Sigma \equiv |\langle e^{-i\mathrm{Im}\,S} \rangle_{\mathrm{Re}\,S} |.
		\end{equation}
		Our simulations employ the well-known HMC algorithm \cite{HMC}, augmented with radial updates and numerical stabilization.
		Radial updates are multiplicative Metropolis-Hastings updates along the radial direction of the non-compact fields.
		They are used to improve convergence properties \cite{kennedy_RU,ostmeyer_RU} and reduce autocorrelation times, especially in the presence of infinite potential barriers \cite{temmen_fully_ergodic}.
		For numerical stability, we employ a procedure based on the separation of scales via QR decompositions when evaluating the fermion determinants and the Green's functions entering the molecular dynamics (MD) evolution \cite{luu_stable}.

		Throughout this section, we fix $\beta = 4$ and $N_t = 32$, and focus on the linear system sizes $L=3,6,9$, corresponding to $N_x = 18,72,162$.
		We set the MD trajectory length to $t_{\mathrm{MD}} = \frac{\pi}{2}\sqrt{V\Delta_t}$ and tune the number of MD steps to obtain acceptance rates of approximately $65\%$.
		The proposal standard deviation of the radial updates is tuned to obtain an acceptance rate of approximately $30\%$.
		For each simulation, we thermalize for $10^3$ steps and record $10^3$ configurations, saving after every $10$th algorithm step.
		We utilize the imaginary field formulation whenever $U\geq KV$ and the real field formulation when $KV>U$. 
		Errors are obtained with bootstrapping.

\begin{figure}
			\centering
			\includegraphics[width=0.55\textwidth]{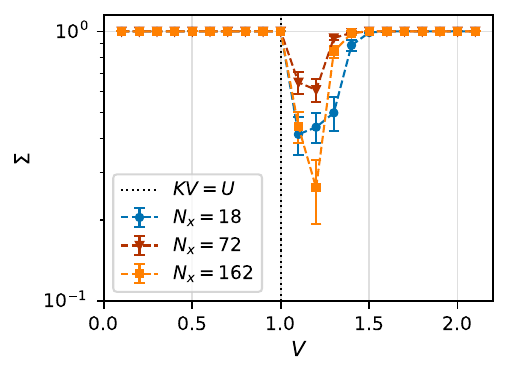}
			\caption{Average phase $\Sigma$ as a function of the nearest-neighbor interaction $V$ at $U = 3$, $\lambda_{\mathrm{SO}} = 0$, $\mu = 0$ and $\beta = 4$, shown for the three system sizes considered. The dotted line marks $KV = U$, where the auxiliary field formulation is switched.}
			\label{fig:V_scan}
		\end{figure}

		We begin at $U=3$, $\lambda_{\mathrm{SO}}=0$, $\mu=0$, and scan over $V\in [0.1, 2.0]$.
		The results are depicted in \figref{fig:V_scan}.
		As anticipated in \secref{sec:two_formulations}, the average phase remains at $\Sigma = 1$ throughout the region $U>KV$.
		Beyond $KV=U$, the fermionic weight loses its positivity and $\Sigma$ develops a pronounced minimum, before returning to unity as $V$ is increased further.
		The minimum deepens from $N_x = 72$ to $N_x = 162$, consistent with the expected exponential decay of $\Sigma$ with volume \cite{troyer_wiese}.
		The smallest lattice does not follow this trend and places the minimum at slightly different $V$, which we attribute to finite-size effects.
		The sign problem encountered in this scan is nonetheless mild and tractable for reweighting, since $\Sigma$ stays above $10^{-1}$ throughout.
		More importantly, $\Sigma$ recovers on both sides of the minimum, so that the limits of small and large $V$ are directly accessible and only a narrow window between them is obstructed.
		Furthermore, the window itself is informative, because an exceptionally severe sign problem has recently been observed to indicate a phase transition, and may be used to locate the critical point \cite{mondaini_quantum_2022,mondaini_universality_2023}.

\begin{figure}
			\centering
			\includegraphics[width=\textwidth]{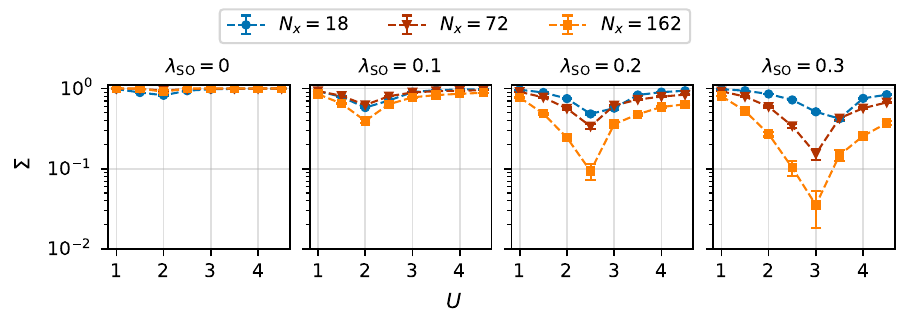}
			\caption{Average phase $\Sigma$ as a function of the on-site interaction $U$ at $KV/U = 1.5$, $\mu = 0$ and $\beta = 4$, shown for the three system sizes considered and for increasing SOC from left to right.}
			\label{fig:U_scan}
		\end{figure}

		We now examine the region $KV>U$ more closely by fixing the ratio $KV/U = 1.5$, scanning over $U\in [1, 4.5]$ and gradually turning on $\lambda_{\mathrm{SO}} \in \{0, 0.1, 0.2, 0.3\}$.
		The results are shown in \figref{fig:U_scan}.
		We again observe a distinct minimum, which deepens and shifts with increasing $\lambda_{\mathrm{SO}}$ and, from $N_x=72$ onwards, also deepens with the system size.
		Under the assumption that the minimum tracks the critical point, this shift implies that the transition moves to larger $U$ as $\lambda_{\mathrm{SO}}$ is increased.
		Overall, the average phase remains above $10^{-1}$ over most of the scanned range and, as in the $V$ scan, recovers toward both small and large $U$.

\begin{figure}
			\centering
			\includegraphics[width=\textwidth]{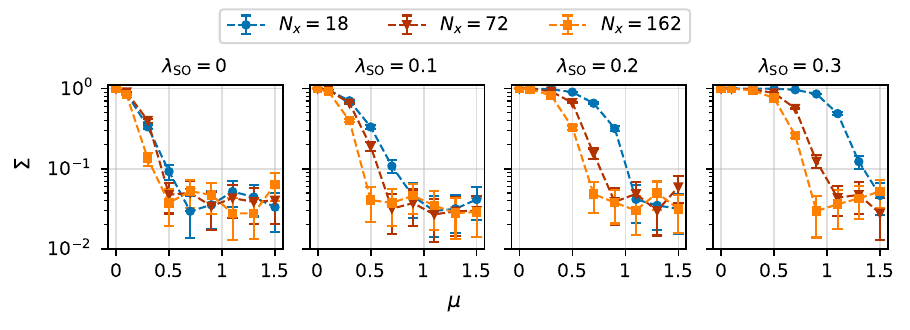}
			\caption{Average phase $\Sigma$ as a function of the chemical potential $\mu$ at $U=2$, $KV/U = 0.9$ and $\beta = 4$, shown for the three system sizes considered and for increasing SOC from left to right.}
			\label{fig:mu_scan}
		\end{figure}

		Finally, we return to the region $U>KV$, fixing $KV/U = 0.9$ and $U=2$.
		We scan over $\mu \in [0, 1.5]$ and again vary $\lambda_{\mathrm{SO}} \in \{0, 0.1, 0.2, 0.3\}$, with the results shown in \figref{fig:mu_scan}.
		As expected, $\Sigma$ is unity at half filling and drops rapidly once $\mu$ becomes finite, quickly rendering simulations infeasible.
		Increasing $\lambda_{\mathrm{SO}}$, however, improves $\Sigma$ at finite $\mu$.
		Read alongside the previous scans, this again suggests a shift of an underlying transition, which is increasingly delayed by the SOC.
		Support for this interpretation comes from the non-interacting limit, where the SOC raises the energy at the Dirac points $\mathrm{K}$ of the Brillouin zone as $E_{\mathrm{K}} = 3\sqrt{3}\lambda_{\mathrm{SO}}$, while leaving the energy $E_{\mathrm{M}} = 1$ at the $\mathrm{M}$ points unchanged.
		The lowest-lying state therefore moves from $\mathrm{K}$ to $\mathrm{M}$ at $\lambda_{\mathrm{SO}} = 1/(3\sqrt{3})\approx0.19$, which lies within the range considered here.

	\section{Summary and outlook}
	\label{sec:summary_outlook}
		In this work, we have developed continuous auxiliary field formulations of the extended Kane-Mele-Hubbard model suitable for simulation with the HMC.
		The freedom in completing the squares of the interaction terms and in the subsequent HS transformation gives rise to several formulations, of which we employ two.
		The first is free of the sign problem at half filling but restricted to dominant on-site interaction, while the second remains applicable once the nonlocal interaction dominates, at the cost of a sign problem.
		Using the average phase as a measure of its severity, we have scanned the parameter space and found the sign problem at half filling to be absent or mild at both dominant on-site and dominant nonlocal interaction, and most severe in between.
		This suggests a route toward the phase diagram, in which the accessible limits are simulated directly and the minimum of the average phase between them is used to locate the transition.

		The results presented here are restricted to small lattices and serve to establish where simulations are possible at all. 
		The next natural step is to extend them to larger volumes, up to $L=18$, and to extract physical observables with a controlled volume extrapolation.
		We intend to do so in both regions, at dominant on-site interaction as a function of the chemical potential and the SOC, and at dominant nonlocal interaction as a function of the on-site interaction and the SOC.
		Moreover, the numerical stabilization employed here allows us to reach considerably larger inverse temperatures than were previously accessible \cite{luu_stable}, which makes it worthwhile to revisit the half-filled case at dominant on-site interaction.

	\begin{acknowledgments}
		This work was funded by the Deutsche Forschungsgemeinschaft (DFG, German Research Foundation) as part of the CRC 1639 NuMeriQS – project no.~511713970.
		The authors gratefully acknowledge the computing time granted through JARA on the supercomputer JURECA~\cite{JURECA} at Forschungszentrum Jülich.
	\end{acknowledgments}

	\bibliographystyle{JHEP}
	\bibliography{ekmh}

\end{document}